\documentclass[useAMS,usenatbib,onecolumn
]{mnras}

\usepackage[usenames,dvipsnames]{xcolor}

\usepackage{amssymb, amsmath}
\usepackage{epsf}
\usepackage{bm}
\usepackage{ulem}

\usepackage{natbib}
\usepackage{graphicx}
\usepackage{cancel}

\def\la{\; \raise0.3ex\hbox{$<$\kern-0.75em\raise-1.1ex\hbox{$\sim$}}\;}
\def\ga{\;  \raise0.3ex\hbox{$>$\kern-0.75em\raise-1.1ex\hbox{$\sim$}}\;}

\title[A note on the ambipolar diffusion in superfluid neutron stars]
{A note on the ambipolar diffusion in superfluid neutron stars}

\author[E. M. Kantor, M. E. Gusakov]
{E.~M.~Kantor  \thanks{kantor@mail.ioffe.ru},
M. E. Gusakov 
\\
Ioffe Physical-Technical Institute of the Russian Academy of
Sciences,
Polytekhnicheskaya 26, 194021 St.-Petersburg, Russia
}

\begin{document}

\date{Accepted 2017 xxxx. Received 2017 xxxx;
in original form 2017 xxxx}

\pagerange{\pageref{firstpage}--\pageref{lastpage}} \pubyear{2017}

\maketitle

\label{firstpage}


\begin{abstract}
We address the problem of magnetic field dissipation in the neutron star cores,
focusing on the role of neutron superfluidity. 
Contrary to the results in the literature, 
we show that in the 
finite-temperature 
superfluid matter composed of neutrons, protons, and electrons, 
magnetic field dissipates exclusively due to 
Ohmic losses and non-equilibrium beta-processes,
and only an 
admixture of muons restores (to some extent) the role 
of particle relative motion for the field dissipation.
The reason for this discrepancy is discussed.
\end{abstract}

\begin{keywords}
stars: neutron -- stars: interiors -- stars: magnetic field
\end{keywords}

\maketitle

\section{Introduction}
\label{Sec:intro}

Since the pioneering works on the magnetic field dissipation 
in the neutron star (NS) cores by \cite{bpp69,hui90,gr92}
a significant progress has been made (e.g., \citealt{su95,us95, td96,us99,hrv08,hrv10,gjs11,bl16,papm17,crv17}),
but many questions have still remained unanswered.
One of these questions regards the effects of nucleon superfluidity and 
superconductivity. 
How do they affect the magnetic field evolution?
In this short note we partly address this question 
by analyzing the effects of neutron superfluidity.
As we 
argue, this analysis is more accurate (and simple), 
than the previous 
treatments (\citealt{gjs11,papm17}), 
which are only valid 
at stellar temperatures $T$ 
much smaller than the neutron critical temperature $T_{cn}$,
and it leads us to interesting conclusions.
In essence, we 
show
that the neutron superfluidity drastically modifies 
the fluid dynamics,
imposing an additional (in comparison to the nonsuperfluid matter) 
constraint on
the velocities of different particle species.
As a result, ambipolar diffusion becomes completely 
irrelevant for the magnetic field evolution 
in the superfluid neutron-proton-electron ($npe$) matter, 
once the stellar temperature $T$ falls (even slightly) 
below 
$T_{cn}$.
Then only Ohmic decay and dissipation due to 
non-equilibrium beta-processes
remain active.
An admixture of muons ($\mu$) introduces additional degree of freedom 
so that dissipation due to particle relative motion can again play a role.

\section{Magnetic field energy dissipation}
\label{Sec:EBdiss}

\subsection{Our assumptions and superfluid equation}
\label{Sec:EBdiss2}

We follow the general strategy outlined in \cite{gr92}.
For simplicity, 
we consider a Newtonian non-rotating NS with the superfluid core composed 
of {\it relativistic} finite-temperature $npe$ (or $npe\mu$) matter, where neutrons are superfluid, 
and protons are normal (nonsuperconducting).
In the absence of magnetic field ${\pmb B}$ the star is in full thermodynamic and hydrostatic equilibrium,
the velocities of all particle species vanish.
We assume that the magnetic field is the only
mechanism that drives the star out of diffusive and beta-equilibrium.
Since the evolution occurs on a very long timescale (\citealt{gr92}),
it proceeds through a set of quasistationary states, 
which means that the time derivatives in the ``Euler equations'' 
(Eqs.\ \ref{nsfl0}, \ref{nsfl}--\ref{psfl} and \ref{esflmu}--\ref{psflmu}, see below), 
as well as in the continuity equations,
can be neglected.
Next, the magnetic field is considered as a small perturbation,
hence the induced particle velocities are also small
and the terms depending on 
square of these velocities
in the dynamic equations can be omitted. 
For simplicity, we also ignore all surface integrals which could appear in the formulas;
they can be easily written out if necessary.

The magnetic energy dissipates when the system evolves towards equilibrium. 
The rate of the magnetic energy change, 
\begin{eqnarray}
\dot{E}_{\rm B}=\int_V \frac{\pmb B}{4\pi}\cdot \dot{\pmb B}\, dV,
\end{eqnarray}
can be presented as (e.g., \citealt{gr92})
\begin{eqnarray}
\dot{E}_{\rm B}=-\int_V \pmb{E}\cdot\pmb{j}\, dV,
\label{EBdiss}
\end{eqnarray}
where ${\pmb E}$ is the electric field; ${\pmb j}$ is the charge current density and
$V$ is the system volume.

Because neutrons are assumed to be in the superfluid state, 
there is an additional degree of freedom in the system -- 
the neutron superfluid velocity ${\pmb V}_{sn}$, 
which is related to the wave function phase $\Phi_n$ of the neutron Cooper-pair condensate 
by the condition ${\pmb V}_{sn} = {\pmb \nabla}\Phi_n/(2 m_n)$. 
Generally, this velocity differs from the velocity of normal neutron component, ${\pmb u}_n$
(i.e., the velocity of neutron thermal Bogoliubov excitations).
At $T<T_{cn}$ both normal and superfluid components contribute
to the neutron current density,
${\pmb j}_n=n_{sn} \pmb {V}_{sn} +(n_n-n_{sn})\pmb{u}_n$,
where $n_{n}$ is the neutron number density and $n_{sn}$ 
is the number density corresponding to superfluid neutron component.
In the absence of rotation
(no Feynman-Onsager vortices)
 ${\pmb V}_{sn}$ obeys the standard ``superfluid'' equation, 
 which is valid at arbitrary $T < T_{cn}$ 
(e.g., \citealt{putterman74,khalatnikov89}; 
\citealt{ga06}; 
note that the quadratically 
small velocity-dependent terms in this equation,
as well as the bulk viscosity terms, 
are already neglected),
\begin{eqnarray}
m_n \frac{\partial \pmb V_{sn}}{\partial t}
+
\pmb{\nabla}\mu_n^\infty=0,
 \label{nsfl0}
\end{eqnarray}
where 
$m_n$ is the neutron bare mass and $\mu_n^\infty$ is the redshifted relativistic neutron chemical potential.
The latter is given by 
$\mu_n^\infty=\mu_n {\rm e}^{\phi/c^2}$, where $\mu_n$ is the neutron chemical potential,
$c$ is the speed of light, 
and $\phi$ is the gravitational potential. 
In a Newtonian star ($\phi \ll c^2$) $\pmb{\nabla} \mu_n^\infty$ can be represented as
$\pmb{\nabla} \mu_n^\infty \approx \pmb{\nabla} \mu_n + \mu_n \pmb{\nabla}\phi/c^2$.

Equation (\ref{nsfl0})  can be further simplified 
by neglecting inertia term which, as we have already discussed above, 
is small for a quasistationary evolving NS. 
Then it reduces to 
\begin{eqnarray}
\pmb{\nabla} \mu_n^\infty=0  \quad \Leftrightarrow \quad
\pmb{\nabla} \mu_n + \mu_n \pmb{\nabla}\phi/c^2=0 \quad {\rm (for \,\, a \,\, Newtonian \,\, star).}
\label{nsfl}
\end{eqnarray}
Equation (\ref{nsfl}), 
which is valid at arbitrary $T< T_{cn}$, 
deserves a comment.
In the NS literature it is customary to use a different form of 
this equation 
(see, e.g., equation 3 in \citealt{gjs11} and equation 1 in \citealt{papm17})
with a friction force density $\pmb{F}_{\rm fr}$ in its right-hand side,
\begin{eqnarray}
\pmb{\nabla}\mu_n^\infty=\frac{\pmb{F}_{\rm fr}}{n_n}.
\label{nsfl2}
\end{eqnarray}
The force density $\pmb{F}_{\rm fr}$ describes 
friction of neutrons with electrons and protons,
and is usually chosen to be equal to
(see, e.g., equations 3, 18, and 40 in \citealt{gjs11} and equations 1, 20, and 21 in \citealt{papm17}),%
%
\footnote{
The expression (\ref{Ffr}) is written for our simplified problem, i.e., assuming 
that protons and electrons are normal (nonsuperconducting).
}
%
%
\begin{equation}
\pmb{F}_{\rm fr}=J_{en} (\pmb{u}_e-\pmb{v}_n)+J_{np}(\pmb{u}_p-\pmb{v}_n),
\label{Ffr}
\end{equation}
where $\pmb{u}_e$ and $\pmb{u}_p$ are the electron and proton  velocities, respectively; 
$\pmb{v}_n\equiv\pmb{j}_n/n_n$ is the velocity of neutron liquid as a whole 
(note that, generally, $\pmb{v}_n\neq \pmb{V}_{sn}\neq \pmb{u}_n$); 
and the `friction' coefficients $J_{en}$ and $J_{np}$ are defined in Sec.\ \ref{Sec:EBdiss3}.

Generally, the `neutron' equation in the form (\ref{nsfl2})
contradicts Eq.\ (\ref{nsfl}).
Both equations coincide only in the limit 
$T\ll T_{cn}$, when $J_{en}$ and $J_{np}$
are suppressed by the neutron superfluidity (e.g., \citealt{gjs11}), 
so that $\pmb{F}_{\rm fr}$ in Eq.\ (\ref{nsfl2}) can be neglected.
So, which equation is correct?

On the one hand, Eq.\ (\ref{nsfl}) is obtained from the 
neutron superfluid equation (\ref{nsfl0}),
which
has a standard form (e.g., \citealt{putterman74}).
%
\footnote{
Note that any substantial modification of this equation 
is forbidden by the basic principles of the theory of superfluidity.
For example, introduction of a friction force in its right-hand side,
\[
m_n \frac{\partial \pmb V_{sn}}{\partial t}
+
\pmb{\nabla}\mu_n^\infty=\frac{\pmb{F}_{\rm fr}}{n_n},
\]
violates the potentiality condition 
for superfluid velocity, 
$\pmb{\nabla}\times {\pmb V}_{sn}=0$,
which must be satisfied in a nonrotating star (e.g., \citealt{khalatnikov89}).
}
%
Note that, this equation remains unchanged 
even for mixtures of nonrotating superfluid
and normal liquids,
as is clearly demonstrated, e.g., in the monograph by \cite{khalatnikov89},
who analyzed dissipative hydrodynamic equations 
for solutions of superfluid helium-II and normal $^3$He
taking into account the diffusion effects
(see Chapter 24 and, in particular, equations 24.37 and 24.30 in that reference).
On the other hand, 
Eq.\ (\ref{nsfl2}) is (as far as we are aware) presented without detailed derivation 
and, as we believe, is the result of unjustified 
application 
of zero-temperature superfluid hydrodynamics
to the case of finite temperatures.
Thus, we conclude, that equation (\ref{nsfl2}) 
is inaccurate at $T \lesssim T_{cn}$ and should be disregarded.

\subsection{$npe$-matter}
\label{Sec:EBdiss3}

The equations of motion for electrons and nonsuperconducting protons take the form
(\citealt{ys91a})
\begin{eqnarray}
-e(\pmb E + \frac{1}{c} \pmb u_e \times \pmb B) 
- \pmb \nabla \mu_e -\frac{\mu_e}{c^2} \, \pmb \nabla \phi
-\frac{J_{ep}}{n_e}(\pmb u_e - \pmb u_p) -\frac{J_{en}}{n_e} (\pmb u_e - \pmb u_n)=0, 
\label{esfl}\\
e (\pmb E + \frac{1}{c} \pmb u_p \times \pmb B) 
-\pmb \nabla  \mu_p -\frac{\mu_p}{c^2}\, \pmb \nabla \phi
-\frac{J_{ep}}{n_p}(\pmb u_p - \pmb u_e) -\frac{J_{np}}{n_p} (\pmb u_p - \pmb u_n)=0, 
\label{psfl}
\end{eqnarray}
where $e$ is the proton electric charge;
$\mu_e$ and $\mu_p$ are the electron and proton 
relativistic chemical potentials, respectively.
Further, 
$n_i$ is the number density of particle species $i=p$, $e$ and
$J_{ik}=J_{ki}$ is the symmetric coefficient
related to the effective relaxation time $\tau_{ik}$ 
for scattering of particles $i$
on particles $k$ by the formula: 
$\tau_{ik}=n_i \mu_i/(c^2 J_{ik})$ 
(see \citealt{ys91a}).   

In Eqs.\ (\ref{esfl}) and (\ref{psfl}) 
thermo-diffusion terms are neglected.

Equations (\ref{nsfl})--(\ref{psfl}) should be supplemented
by the total force balance equation,
which, for the problem in question, takes the standard form
(the same for superfluid and nonsuperfluid liquids),
\begin{eqnarray}
\frac{\pmb j \times \pmb B}{c}=\pmb \nabla P+\frac{(P+\epsilon)}{c^2} 
\pmb \nabla \phi= \sum_{i=n,p,e} \left(n_i \pmb \nabla \mu_i+n_i \frac{\mu_i}{c^2} \, \pmb \nabla \phi \right), 
\label{euler}
\end{eqnarray}
where $P$ and $\epsilon$ are the pressure and energy density, 
and $\pmb j=e n_p {\pmb u}_p - e n_e {\pmb u}_e$. 
Let us now compose the following combination, 
$[n_e\times$ (\ref{esfl}) $+n_p\times$ (\ref{psfl})$-n_n\times$ (\ref{nsfl})]$-$(\ref{euler}). 
Taking into account the quasineutrality condition, $n_e=n_p$, we get
\begin{eqnarray}
J_{en} (\pmb u_e - \pmb u_n)+J_{np} (\pmb u_p - \pmb u_n)=0.  
\label{velrel}
\end{eqnarray}
This equation imposes an additional 
(in comparison to the non-superfluid matter) constraint on the velocities ${\pmb u}_i$. 
For example, if we neglect collisions between 
neutrons and electrons ($J_{en}=0$), we obtain $\pmb u_p=\pmb u_n$.

Now, summing up $n_e\times$ (\ref{esfl})$+n_p\times$ (\ref{psfl})$+n_e\times$ (\ref{nsfl}), 
we arrive at
\begin{eqnarray}
\frac{\pmb j \times \pmb B}{c}=-n_p \pmb \nabla \Delta \mu_e-n_p \, \frac{\Delta \mu_e}{c^2}\,  \pmb \nabla \phi= -{\rm e}^{-\phi/c^2} n_p \pmb \nabla (\Delta \mu_e \;{\rm e}^{\phi/c^2}),
\end{eqnarray}
where $\Delta \mu_e \equiv  \mu_n- \mu_p- \mu_e$. 
In the Newtonian limit $n_p (\Delta \mu_e/c^2) 
\nabla \phi \ll n_p \nabla \Delta \mu_e$, and we have
\begin{eqnarray}
\frac{\pmb j \times \pmb B}{c}=-n_p \pmb \nabla \Delta \mu_e.  \label{jB}
\end{eqnarray}
This is a Grad-Shafranov type equation 
for the magnetic field, as in the case of magnetic equilibria in barotropic fluids.
Note that the Lorentz force density $\pmb j \times \pmb B/c$ 
depends on 
the gradient of only one scalar function 
(in contrast to the non-superfluid matter, 
where it 
depends on
the gradients of two scalar functions, see \citealt{gl16}). 
This means that only very specific magnetic field configurations can 
restore
hydrostatic equilibrium when neutrons are superfluid (\citealt{gl16}). 
Thus, once NS temperature drops below the neutron critical temperature $T_{cn}$ 
at a given point, 
the magnetic field has to rearrange itself 
to meet the new hydrostatic equilibrium condition (\ref{jB}). 
This rearrangement may result in magnetar
activity and effective dissipation
of the magnetic field energy
on a typical timescale of NS cooling.%
%
\footnote{As results of \cite{gl16} indicate, the same conclusion also applies if protons 
are superconducting.}
%

To find the dissipation rate $\dot{E}_B$ (\ref{EBdiss}),
we express $\pmb E$ from  Eq.\ (\ref{psfl}) for protons,
\begin{eqnarray}
\pmb E=- \frac{\pmb u_p\times \pmb B}{c}+\frac{\pmb \nabla \mu_p}{e}+\frac{\mu_p \pmb \nabla \phi}{c^2 e}+\frac{ J_{ep} (\pmb u_p-\pmb u_e) + J_{np} (\pmb u_p-\pmb u_n)}{ e n_p}. 
\label{EE}
\end{eqnarray}
The second and third
terms in Eq. (\ref{EE}) 
are potential %
%
\footnote{
In the approximation of a Newtonian star ($\phi/c^2\ll 1$), employed in this paper, 
these terms can be presented as
$\pmb{\nabla} \mu_p/e+\mu_p \pmb{\nabla} \phi/(c^2 e)=
{\rm e}^{-\phi/c^2}\pmb{\nabla}(\mu_p \, {\rm e}^{\phi/c^2}/e)
\approx \pmb{\nabla}(\mu_p \, {\rm e}^{\phi/c^2}/e)$  
and hence are indeed potential.
It is interesting to note that fully relativistic calculation would not change this result.
}
%
and thus do not contribute 
to the magnetic field dissipation 
(to see this, integrate Eq.\ \ref{EBdiss} by parts
and use the continuity equation, $\pmb{\nabla } \cdot {\pmb j}=0$). 
Thus,
\begin{eqnarray}
\dot{E}_{\rm B}=-\int_V \left[- \frac{\pmb u_p\times \pmb B}{c}+\frac{  J_{ep} (\pmb u_p-\pmb u_e) + J_{np} (\pmb u_p-\pmb u_n) }{ e n_p}\right]\cdot \pmb{j}\, dV.
\label{Eb2}
\end{eqnarray}
The first term here can be rewritten as
\begin{eqnarray}
\int_V \left(\frac{\pmb u_p\times \pmb B}{c}\right)\cdot \pmb{j}\, dV=-\int_V \left(\frac{\pmb j\times \pmb B}{c}\right)\cdot\pmb{u}_p\, dV.
\label{13}
\end{eqnarray}
Substituting now (\ref{jB}) into (\ref{13})
\begin{eqnarray}
-\int_V \left(\frac{\pmb j\times \pmb B}{c}\right)\cdot \pmb{u}_p \, dV
=\int_V \left(n_p \pmb \nabla \Delta \mu_e \right)\cdot \pmb{u}_p\, dV,
\end{eqnarray}
and integrating by parts, one finds (the integral over the distant surface is omitted)
\begin{eqnarray}
\int_V \left(\frac{\pmb u_p\times \pmb B}{c}\right)\cdot \pmb{j}\, dV=
\int_V - \pmb{\nabla}\cdot (n_p \pmb{u}_p) \Delta \mu_e \, dV.
\label{15}
\end{eqnarray}
The divergence term here can be expressed with the help of the
continuity equation for protons, $\pmb{\nabla} \cdot (n_p \pmb{u}_p)=\Delta \Gamma$, 
where the source $\Delta \Gamma$ 
accounts for the non-equilibrium beta-processes.
When $\Delta \mu_e \ll k_{\rm B} T$, 
$\Delta \Gamma$ can be approximated as
$\Delta \Gamma\approx \lambda_e \Delta \mu_e$ 
($\lambda_e$ is the density and temperature-dependent coefficient
and $k_{\rm B}$ is the Boltzmann constant),
so that (\ref{15}) reduces to
\begin{eqnarray}
\int_V \left(\frac{\pmb u_p\times \pmb B}{c}\right)\cdot \pmb{j}\, dV=
\int_V -\lambda_e \Delta \mu_e^2   \, dV.
\end{eqnarray}
Returning now to Eq.\ (\ref{Eb2}), it can be represented as
\begin{eqnarray}
\dot{E}_{\rm B}=-\int_V \pmb{E}\cdot \pmb{j}\, dV=
\int_V \left[-\lambda_e \Delta \mu_e^2 -J_{en} (\pmb u_e - \pmb u_n)^2 - J_{ep} (\pmb u_e - \pmb u_p)^2 - J_{np} (\pmb u_n - \pmb u_p)^2 \right]\, dV \nonumber \\
+\int_V (\pmb u_e - \pmb u_n) \left[J_{en} (\pmb u_e - \pmb u_n) + J_{np} (\pmb u_p - \pmb u_n)\right] \, dV,
\end{eqnarray}
or, in view of Eq. (\ref{velrel}),
\begin{eqnarray}
\dot{E}_{\rm B}=-\int_V \pmb{E} \cdot \pmb{j}\, dV=
\int_V \left[-\lambda_e \Delta \mu_e^2 -J_{en} (\pmb u_e - \pmb u_n)^2 - J_{ep} (\pmb u_e - \pmb u_p)^2 - J_{np} (\pmb u_n - \pmb u_p)^2 \right]\, dV. 
\label{EBdissnpe}
\end{eqnarray}
Clearly, magnetic field dissipates
due to particle mutual transformations and relative motion (diffusion). 
In principle, the same expression can also be derived 
for non-superfluid $npe$ matter (Gusakov et al., in preparation), 
but now the velocities are related by the constraint (\ref{velrel}). 
Using this constraint, Eq.\ (\ref{EBdissnpe}) can be rewritten as
\begin{eqnarray}
\dot{E}_{\rm B}=
\int_V \left[-\lambda_e \Delta \mu_e^2  - \frac{j^2}{\sigma_0} \right]\, dV, 
\end{eqnarray}
where $\sigma_0=e^2 n_e^2/(J_{ep}+\frac{J_{en}J_{pn}}{J_{en}+J_{pn}})$ 
is the electrical conductivity in the absence of a magnetic field.
We come to an important conclusion that
in superfluid $npe$ matter there is no magnetic field dissipation 
due to ambipolar diffusion:
Magnetic field dissipates exclusively 
due to Ohmic decay and non-equilibrium 
particle 
transformations. 
The former is extremely inefficient 
in neutron stars%
%
\footnote{A typical timescale is 
$t_{\rm Ohmic}=4\pi L^2 \sigma_0/c^2 \sim 10^{14}T_8^{-5/3}\, \rm yrs$, 
where $L$ is the lengthscale of the magnetic field variation
(we take $L \sim 10^6 \,\rm cm$)
and $T_8$ is the temperature of the NS core, normalized to $10^8\,\rm K$ (\citealt{shternin08}).}, 
%
while the latter strongly depends 
on the rate of non-equilibrium beta-processes in NS matter.
A typical dissipation timescale associated with these processes can be estimated as
$t_{\rm reactions} \sim B^2/(4 \pi \lambda_e \delta \mu_e^2) \sim 4 \pi n_p^2/(\lambda_e B^2)$.
In the case of modified Urca (mUrca) processes
this estimate gives%
%
\footnote{The coefficient $\lambda_e$ is estimated using 
the formulas given in the review by \cite{ykgh01}.}
%
$t_{\rm reactions}\gtrsim 3 \times 10^{12} \,  T_8^{-6}B_{14}^{-2}\,\rm yrs$, 
too much to affect the magnetic field evolution.
In turn, for 
the direct Urca (dUrca) process we get  
$t_{\rm reactions}\gtrsim 2\times 10^4 \, T_8^{-4}B_{14}^{-2}\, \rm yrs$, 
i.e., dUrca can be effective dissipation agent 
for sufficiently strong magnetic fields.

It should be emphasized that the fact that magnetic field does not dissipate
through the ambipolar diffusion in the NS region where neutrons are superfluid,
was clearly realised long  ago by \cite{us95,us99}.
However, these authors only considered the case of vanishing temperature ($T=0$),
when all neutrons condense in Cooper pairs and simply cannot scatter off the protons.
In contrast, here we argue that ambipolar diffusion is not important 
for {\it any} temperature, even slightly smaller than the critical temperature $T_{cn}$
(when almost all neutrons are unpaired).
Note that, this conclusion is in contrast to the generally held view 
(e.g., \citealt{gjs11,papm17})
about the possible important role of ambipolar diffusion at temperatures $T$ comparable to $T_{cn}$,
which is based on the analysis of equations, strictly  valid only at $T \ll T_{cn}$ 
(see Sec.\ \ref{Sec:EBdiss2} for details).

\subsection{$npe\mu$-matter}
\label{Sec:EBdiss4}

An admixture of muons introduces an additional degree of freedom into the system.
The dynamic equations for superfluid $npe\mu$ mixture consist of 
superfluid (neutron) Eq.\ (\ref{nsfl}) together with the three equations
on the charged components,
\begin{eqnarray}
-e(\pmb E + \frac{1}{c} \pmb u_e \times \pmb B) - \pmb \nabla \mu_e -\frac{\mu_e}{c^2} \pmb \nabla \phi-\frac{J_{ep}}{n_e}(\pmb u_e - \pmb u_p) -\frac{J_{en}}{n_e} (\pmb u_e - \pmb u_n)-\frac{J_{e\mu}}{n_e} (\pmb u_e - \pmb u_\mu)=0, \label{esflmu}\\
-e(\pmb E + \frac{1}{c} \pmb u_\mu \times \pmb B) - \pmb \nabla \mu_\mu -\frac{\mu_\mu}{c^2} \pmb \nabla \phi-\frac{J_{\mu p}}{n_\mu}(\pmb u_\mu - \pmb u_p) -\frac{J_{\mu n}}{n_\mu} (\pmb u_\mu - \pmb u_n)-\frac{J_{e\mu}}{n_\mu} (\pmb u_\mu - \pmb u_e)=0, \label{musflmu}\\
e (\pmb E + \frac{1}{c} \pmb u_p \times \pmb B) -\pmb \nabla  \mu_p -\frac{\mu_p}{c^2} \pmb \nabla \phi-\frac{J_{ep}}{n_p}(\pmb u_p - \pmb u_e) -\frac{J_{np}}{n_p} (\pmb u_p - \pmb u_n)-\frac{J_{\mu p}}{n_p} (\pmb u_p - \pmb u_\mu)=0. \label{psflmu}
\end{eqnarray}
In Eqs.\ (\ref{esflmu})--(\ref{psflmu}) 
$\pmb u_\mu$ and $\mu_\mu$ are the muon velocity 
and relativistic chemical potential, respectively.
In analogy with Eq.\ (\ref{jB}) 
one can derive the force balance equation 
for
superfluid $npe\mu$ matter,
\begin{eqnarray}
\frac{\pmb j \times \pmb B}{c}=-n_e \pmb \nabla \Delta \mu_e-n_\mu \pmb \nabla \Delta \mu_\mu,
\label{eqmu}
\end{eqnarray}
and find that the Lorentz force density is determined by the gradients of two scalars, 
$\pmb \nabla \Delta \mu_e$ and $\pmb \nabla \Delta \mu_\mu$, 
where
$\Delta \mu_\mu \equiv  \mu_n- \mu_p- \mu_\mu$. 
Therefore, comparing to superfluid $npe$-matter, 
there is more freedom to choose possible magnetic field configurations.
Proceeding in a similar way as in the case of $npe$ matter, 
one can show that 
the magnetic field dissipation rate is given by %
%
\footnote{The same expression is also valid for non-superfluid $npe\mu$ matter,
but then it is not constrained by Eq.\ (\ref{velrelmu}).}
%
\begin{eqnarray}
\dot{E}_{\rm B}=
\int_V \left[-\lambda_e \Delta \mu_e^2-\lambda_\mu \Delta \mu_\mu^2 -\frac{1}{2}\sum_{i,k,i\neq k}J_{ik} (\pmb u_i - \pmb u_k)^2 \right]\, dV, \
\label{EBdissnpemu}
\end{eqnarray}
while the velocities ${\pmb u}_i$ 
are related by (compare this result with the constraint \ref{velrel})
\begin{eqnarray}
J_{\mu n} (\pmb u_\mu - \pmb u_n)+J_{en} (\pmb u_e - \pmb u_n)+J_{np} (\pmb u_p - \pmb u_n)=0. 
\label{velrelmu}
\end{eqnarray}
In Eq.\ (\ref{EBdissnpemu}) the indices $i,k$ run over $n,p,e,\mu$;
$\lambda_\mu$ has the same meaning as $\lambda_e$,
but is defined for Urca-reactions involving muons.
Generally, as follows from Eqs.\ (\ref{EBdissnpemu}) and (\ref{velrelmu}),
the magnetic field dissipation in $npe\mu$ matter 
is {\it not}
associated exclusively
with
the Ohmic decay and non-equilibrium beta-processes.
Even if we neglect the lepton interactions with neutrons 
($J_{en}=J_{\mu n}=0$), and
find that $\pmb u_p =\pmb u_n$ from Eq.\ (\ref{velrelmu})
(i.e., normal neutrons move with protons), 
the relative motion between electrons and muons will still enter the dissipation rate.
The effect of such motion on the magnetic field dissipation 
can be relatively small, especially at large densities 
(when muons more and more resemble electrons). But
this should be checked by direct calculation which is beyond the scope
of the present short note.
Concluding, the presence of muons (or other, more exotic, particle species)
complicates the things and may, in principle, 
affect the magnetic field evolution in superfluid NSs.

\section{Conclusions}
\label{Sec:Conclusions}

In this note we considered a simplified illustrative problem 
of the magnetic field evolution 
in a NS, whose
core is composed of superfluid neutrons, 
nonsuperconducting protons and electrons 
with possible admixture of muons.  
A star is assumed to be non-rotating, i.e., it does not have neutron 
vortices in its interiors.
Perturbation of the system by the magnetic field is assumed to be small 
and the NS evolution is supposed to be quasistationary -- 
standard assumptions (e.g., \citealt{gr92}), that allow us to neglect time derivatives and velocity-dependent nonlinear terms 
in the Euler-type and particle continuity equations.
Bearing in mind simplifications described above, 
we arrived at the following conclusions:

1. Ambipolar diffusion is irrelevant for the magnetic field dissipation 
in superfluid $npe$ matter at temperatures $T$ even slightly smaller than $T_{cn}$. 
Magnetic field in this case dissipates only
because of
Ohmic losses 
(inefficient mechanism in the neutron star cores) 
and 
non-equilibrium Urca processes
(can be efficient if the direct Urca process is open).
This result is in contrast to the results of \cite{gjs11} and \cite{papm17}
who, as we argue in Sec.\ \ref{Sec:EBdiss2}, 
used a superfluid dynamic equation for neutrons, which 
is correct only in the limit $T\ll T_{cn}$. 

2. Since only very specific magnetic field configurations 
can support
hydrostatic equilibrium 
in superfluid $npe$ matter 
(see section \ref{Sec:EBdiss} and the work by \citealt{gl16}), 
magnetic field should 
``feel'' expansion of the superfluid region upon NS cooling
and reorganize itself accordingly
on a cooling timescale.
This may result in an increased magnetic activity, e.g., in magnetars.

3. An admixture of muons will restore, to some extent,
the role of ambipolar diffusion for the magnetic field evolution, 
although the 
magnetic energy dissipation rate (\ref{EBdissnpemu})
will differ from that in non-superfluid matter because of 
($i$) suppression of $np$, $ne$, and $n\mu$ collisions by neutron superfluidity (e.g., \citealt{ys91b})
and ($ii$) an additional constraint (\ref{velrelmu}) 
relating normal particle velocities ${\pmb u}_i$ ($i=n$, $p$, $e$, $\mu$).
In particular, if we neglect 
the lepton-neutron collisions,
there will be no relative motion between normal neutrons and protons, 
${\pmb u}_n={\pmb u}_p$.

\section{Acknowledgements}

This study was supported by the Russian Science Foundation 
(grant number 14-12-00316).



\label{lastpage}

\end{document}